\documentstyle[aps]{revtex}
\begin{document}
\draft
\title{Theory and Applications of the Systematic Detection of Unstable
  Periodic Orbits in Dynamical Systems}\author{Detlef Pingel and Peter Schmelcher}
\address{Theoretical Chemistry, Institute for Physical Chemistry, INF 229,\\
University of Heidelberg, 69120 Heidelberg, Germany }
\author{Fotis K. Diakonos}
\address{Department of Physics, University of Athens,\\ GR-15771 Athens, Greece}
\author{Ofer Biham}
\address{Racah Institute of Physics, The Hebrew University,\\ Jerusalem 91904, Israel}
\date{\today}
\maketitle
\begin{abstract}
  A topological approach and understanding to the detection of
  unstable periodic orbits based on a recently proposed method (PRL
  {\bf 78}, 4733 (1997)) is developed. This approach provides a
  classification of the set of transformations necessary for the
  finding of the orbits.  Applications to the Ikeda and H\'enon map
  are performed, allowing a study of the distributions of Lyapunov
  exponents for high periods. Particularly the properties of the least
  unstable orbits up to period 36 are investigated and discussed.
\end{abstract}

\section{Introduction}

Unstable periodic orbits (UPOs) represent a skeleton of complex
chaotic systems and allow the calculation of many characteristic
quantities of the underlying dynamics like Lyapunov exponents, fractal
dimensions and entropies of the attractors
\cite{Grebogi88,Cvitanovic88a}.  For dissipative systems expansions in
terms of periodic orbits are well established in the literature
\cite{Grebogi88,Auerbach87,Cvitanovic88b,Ott89,Lathrop89} and
demonstrate the relevance of the cycles for the understanding of
chaotic dynamics.  Both low-dimensional model systems such as discrete
maps \cite{Grebogi88,Cvitanovic88a} as well as experimental time
series \cite{Schwartz92,Abarbanel93,Badii94,So96} have been studied.
Furthermore the series expansion of semiclassical
properties of classically chaotic Hamiltonian systems with respect to
the length and stability coefficients of the periodic orbits is a
fruitful and frequently applied technique: it allows the investigation
of the energy level density as well as other quantum properties
\cite{Gutzwiller90}.  Much has been added to the importance of the
UPOs by using them to control chaotic dynamical systems ( see
\cite{Ott93} and references therein ).

More recently, cycle expansion techniques have been invented and shown
to converge well, especially when the symbolic dynamics is well
understood \cite{Cvitanovic88b,Artuso90}.  Series expansions over
periodic orbits used for calculating dynamical averages are typically
ordered according to the orbit length $p$
\cite{Grebogi88,Artuso90,Grassberger89,Biham92}.  Drawbacks of these
expansions are the large number of orbits (increasing exponentially
with $p$), the required completeness of the set of cycles for given
period and the slow convergence properties
\cite{Grassberger89,Biham92}.  A promising proposal was made
\cite{Dettmann97a,Dettmann97b} saying that series expansions could
converge better if they are truncated according to the stability of
the orbits \cite{Dahlqvist91}.  What is more, stability ordering does
not rely on the knowledge of the systems symbolic dynamics, which is
unknown for generic dynamical systems.

Chaotic dynamics is intrinsic for many physical systems, which is why
periodic orbit theory is not restricted to specific areas of physics.
Much effort has been spent on developing efficient techniques for
calculating UPOs of a given dynamical system.  What makes this quest
so difficult is the exponential proliferation of the number of cycles
with increasing period and their increasing instability (in fully
chaotic systems).  Specifically the modified Newton--Raphson method is
an all--purpose method, which does not require a special form of the
underlying equations of motion. However, it needs a good initial guess
for the starting points. Therefore this algorithm rapidly becomes
expensive and is limited to relatively short periods and
low--dimensional systems.  A variety of other methods have been
developed which focus either on time series analysis or on finding
UPOs for given equations of motions
\cite{Cvitanovic88a,Cvitanovic88b,Grassberger89,Han95,Zol98,Biham89}.
For a special class of
systems, a numerical technique for calculating arbitrarily long UPOs
to any desired accuracy was introduced in \cite{Biham89} for the
H\'enon map \cite{Henon76} and later applied to a few other dynamical
systems \cite{Biham92,Biham90,Wenzel91}.  This method allows the
systematic computation of all UPOs of any given order, each given by a
unique binary symbol sequence.

Recently, we proposed an alternative method (in the following referred
to as SD method) for the calculation of UPOs \cite{Schmelcher97}.  The
basic idea is to transform the fully chaotic system to a new dynamical
system with the periodic orbits keeping their positions but changing
their stability properties: For a particular transformed system a
certain fraction of the periodic cycles becomes stable and can be
found by simply iterating the transformed system.  This fraction
depends on a tuning parameter which represents an upper stability
cutoff for the fixed points (FPs) to be detected.  In the following we
will use the term 'stabilisation' and 'stabilised fixed points' for
the process and fact respectively that the unstable FPs of the
original chaotic dynamical system have become stable in the
corresponding transformed system. The reader is kindly asked to
distinguish this use of the term 'stabilisation' from the one used in
control theory of chaos via unstable periodic orbits.  According to
the above the SD method allows the systematic calculation of the least
unstable periodic orbits of any given order $p$ \cite{Diakonos98}.
The latter possibility meets the requirements of the series expansions
using stability ordering, since they allow to derive properties of a
physical system by exploring only the least unstable period orbits up
to a given stability cutoff.  The technique of the SD method is highly
flexible and can be applied in a straightforward manner to a great
variety of discrete dynamical systems of any dimension.  Continuous
dynamical systems can thereby be treated using properly chosen
Poincar\'e surfaces of section.  Therefore, it might open the way for
employing the proposal of \cite{Dettmann97a,Dettmann97b} for a great
variety of dynamical systems.

The present work has a twofold aim. The first goal is to enhance our
understanding of and give insights into the SD method. To that end we
provide a complete classification of the involved transformations.  As
a result we gain both topological as well as geometrical understanding
and interpretation of the transformations. Corresponding invariant
structures are thereby revealed and the FPs can be classified
similarly to the stabilising transformations. This opens the future
perspective to selectively detect UPOs not 'only' with respect to
their stability but also with respect to certain desired geometrical
properties. We will thereby learn how simple global operations on
the dynamical system change the stability properties of fixed points. 
The second aim of this work is to elucidate and extend the work in
ref. \cite{Schmelcher97}.  To this end we provide extensive results of
applications of our approach to the Ikeda \cite{Ikeda79} and H\'enon \cite{Henon76}
maps.
For the Ikeda map, we calculate the complete sets of FPs for periods
up to $p=15$.  The number of orbits is large enough to investigate the
distribution of Lyapunov exponents. This distribution is compared with
the corresponding one of the H\'enon map according to
\cite{Biham89}.  Next the algorithm of \cite{Diakonos98} to determine
the least unstable orbits is slightly modified and applied to the
Ikeda and H\'enon map.  The ten most stable orbits of a given length
up to period $p=36$ are studied for both maps.  Their Lyapunov
exponents vary in a characteristic way as a function of the period.

In detail we proceed as follows: In Section 2.1 we briefly review the
SD method for finding UPOs as described in \cite{Schmelcher97}.  The
topological/geometrical classification and geometrical extension is
presented and discussed in Sec. 2.2.  In Sec. 3.1 we show the results
of complete sets of orbits and we present and analyse the distribution
of Lyapunov exponents in Sec. 3.2.  The extended method to stabilise
the least unstable cycles is given in Sec. 3.3 together with the
corresponding results for the H\'enon and Ikeda map.  Sec. 4 concludes
with a summary.

\section{Theory of the Stabilisation of Fixed Points}
\subsection{Brief Review of the underlying Method}

In order to be self--contained and for our further theoretical
investigation we recall in the following the key ideas of the method
developed in ref. \cite{Schmelcher97} to detect UPOs. Consider a
discrete chaotic dynamical system
$U:~{\mathbf{r}}_{i+1}=\vec{f}({\mathbf{r}}_i)$ in $n$ dimensions.  The FPs of
the $p$th iterate $\vec{f}^{(p)}$ are points of the UPOs of period
$p$.  To find the FPs of $U$ the following strategy is employed: A set
of transformations is specified which transforms $U$ into new
dynamical systems $\{S_{k \sigma}\}$ {\sl with the FPs keeping their
  original locations in space}. The set $\{S_{k \sigma}\}$ is chosen
such that for each unstable FP $\alpha_u$ of $U$ there exists a
specific transformed system $S_{k^\prime \sigma^\prime}$ of the set
$\{S_{k \sigma}\}$ for which this FP has become dissipatively stable
(i.e. $\mbox{Re}(\lambda_i)<0$ for continuous systems, $|\lambda_i|<1$
for discrete systems, with eigenvalues $\lambda_i$) and can therefore
be detected by simply iterating properly chosen starting points of the
transformed system $S_{k^\prime \sigma^\prime}$.  For each
$S_{k^\prime \sigma^\prime}$ a different set of FPs of $U$ is
stabilised.  Let us denote by $I_{\min}$ the minimal set of pairs $(k
\sigma)$ with the property that there exists for each unstable FP
$\alpha_u$ at least one pair $(k^\prime \sigma^\prime) \in I_{\min}$
for which $S_{k^\prime \sigma^\prime}$ transforms $\alpha_u$ into a
stable FP $\alpha_s$.  This set holds for arbitrary period $p$. The
search for the FPs of $U$ is then straightforward: A starting point
chosen in the global neighbourhood (see below) of the FP $\alpha_u$
iterated with the transformed dynamical system $S_{k^\prime
  \sigma^\prime}$, $(k^\prime \sigma^\prime) \in I_{\min}$ converges,
due to the stability of $\alpha_s$, to the position of $\alpha_s$
which is equal to that of $\alpha_u$.

Propagating a set $\{{\mathbf{r}}_i\}$ of starting points and using all $(k
\sigma)\in I_{\min}$ we end up with a set of FP of $U$ whose
completeness can be ensured by successively enlarging the set
$\{{\mathbf{r}}_i\}$.  Let us now specify the systems $S_{k \sigma}$:
\begin{equation}
S_{k \sigma}:~~{\mathbf{r}}_{i+1}={\mathbf{r}}_i+{\mathbf{\Lambda}}_{k \sigma} 
[\vec{f}({\mathbf{r}}_i) - {\mathbf{r}}_i]
\label{dynsyss}
\end{equation}
\noindent
$S_{k \sigma}$ are linear transformations of the original dynamical
law $U$.  ${\mathbf{\Lambda}}_{k \sigma}$ are invertible constant $n
\times n$ matrices.  The definition of $S_{k \sigma}$ and $U$ clearly
shows that their FPs are one to one and at the same positions.  Eqn.
(\ref{dynsyss}) implies the following relation for the stability
matrices ${\mathbf{T}}_U$ and ${\mathbf{T}}_{S_{k \sigma}}$ of $U$ and
$S_{k \sigma}$, respectively:
\begin{equation}
{\mathbf{T}}_{S_{k \sigma}}=\mathbf{1}+{\mathbf{\Lambda}}_{k \sigma}({\mathbf{T}}_U-\mathbf{1})\label{stamattrafo}
\end{equation}
In ref. \cite{Schmelcher97} it was shown that ${\mathbf{\Lambda}}_{k
  \sigma}$ can be cast in the form ${\mathbf{\Lambda}}_k=\lambda\cdot
{\mathbf{C}}_{k \sigma}$ with $0 < \lambda < 1$.  The set of matrices
$\{{\mathbf{C}}_{k \sigma}\}$ contains all orthogonal matrices with only
one non-vanishing entry $\pm 1$ per row or column, i.e. they represent
a group of special reflections and permutations.

$\sigma=\pm$ indicates the sign of the matrix determinant, and $k$ is
an additional label to uniquely specify the matrices. In two dimensions we have
${\mathbf{C}}_{0+}=-{\mathbf{C}}_{2+}=\left({1\atop0}{0\atop 1}\right)$,
${\mathbf{C}}_{0-}=-{\mathbf{C}}_{2-}=\left({-1\atop 0}{0\atop1}\right)$,
${\mathbf{C}}_{1+}=-{\mathbf{C}}_{3+}=\left({0\atop -1}{1\atop 0}\right)$
and ${\mathbf{C}}_{1-}=-{\mathbf{C}}_{3-}=\left({0\atop -1}{-1\atop
    0}\right)$.  The matrices $\{{\mathbf{C}}_{k
  \sigma}|k=0...3;\sigma=\pm\}$ form a group with
$\{{\mathbf{C}}_{k+}|k=0,...,3\}$ (matrices with positive determinant)
being a subgroup of order 4. Table \ref{cmult} is the corresponding
multiplication table.  Obviously, the product of two matrices is
\begin{equation}
{\mathbf{C}}_{k^{\prime\prime} \sigma^{\prime\prime}}=
{\mathbf{C}}_{k \sigma}\cdot{\mathbf{C}}_{k^\prime \sigma^\prime}
~~\mbox{with}~~ k^{\prime\prime}=k^\prime+ \sigma^\prime k \bmod 4
~~\mbox{and}~~ \sigma^{\prime\prime}=\sigma\sigma^\prime
\label{cmulteqn}
\end{equation}
The notation introduced above is 
different from the one used in ref. \cite{Schmelcher97} and will reveal its meaning in the 
course of the 
classification of all possible FPs (occurring in the original and
transformed system), as provided later on. 
The arithmetics with respect to the first index ($k,k^\prime,k^{\prime\prime},..$)
has always to be taken modulo 4.
We remark that the minimal set $I_{\min}$ is significantly smaller
than the set of pairs $(k \sigma)$ belonging to the matrices $\{S_{k
  \sigma}\}$.  Given a certain unstable FP $\alpha_u$ the above choice
of linear transformations represented by the set $\{{\mathbf{C}}_{k
  \sigma}\}$ of matrices allows to find a particular
${\mathbf{C}}_{k^\prime \sigma^\prime}$ such that the real parts of the
eigenvalues of ${\mathbf{C}}_{k^\prime
  \sigma^\prime}({\mathbf{T}}_{U}-\mathbf{1})$
are negative at the position of the FP. As a consequence (see Eqn.
(\ref{stamattrafo})), if $\lambda$ is chosen sufficiently small, the
magnitude of the real parts of the eigenvalues of the FP $\alpha_s$ in
the transformed system are smaller than one and we therefore encounter
a stable FP which can be detected as described above.
The critical value of $\lambda$, which just suffices to make the FP
stable, can immediately be read off from the quadratic equations
relating the stability coefficients of the original system $U$ and
those of the transformed system $S_{k \sigma}$ \cite{Schmelcher97}.
The above procedure can easily be extended to higher iterates
$\vec{f}^{(p)}({\mathbf{r}})$ of $U$ (by replacing $\vec{f}$ with
$\vec{f}^{(p)}$ in Eq. (\ref{dynsyss})) allowing us to determine all
order $p$ cycles of $U$.

The advantage of the SD method is clearly its global character in the sense that
even points far from the linear neighbourhood of a FP are attracted
close to the FP after a finite number of iterations of the transformed
dynamical law.  The basin of attraction of a single stabilised FP is a
simply connected area in phase space.  The typical number of starting
points needed to obtain the UPOs of a given length on the attractor is
only slightly more than the expected number of cycle points
themselves.

The parameter $\lambda$ is a key quantity here.  It is related to the
stability of the desired cycle in the transformed system.  With
increasing period of the cycles, $\lambda$ has to be reduced to
achieve stabilisation of all FPs.  One may, however, also be
interested in the most stable periodic orbits of a given period $p$
\cite{Diakonos98} which is one of the key issues of the present work.
In this context $\lambda$ operates as a filter allowing the selective
stabilisation of only those UPOs which possess Lyapunov exponents
smaller than a critical value.  Therefore, starting the search for
UPOs within a certain period $p$ with a value $\lambda \cong
O(10^{-1})$ and gradually lowering $\lambda$ we obtain the sequence of
all unstable orbits of order $p$ sorted with increasing values of
their Lyapunov exponents.  In \cite{Diakonos98} it was shown that for
the specific choice $(k \sigma) =(0+) \in I_{\min}$ the relation
between $\lambda$ and the stability coefficients of the FPs of the
original system $U$ is a strict monotonous one.  Transformed dynamical
systems $S_{k \sigma}$ belonging to other pairs $(k^\prime
\sigma^\prime) \not= (0+)$ do not obey such a strict behaviour but
show a rough ordering of the sequence of stability eigenvalues of the
FPs of $U$ stabilised in the course of decreasing values for
$\lambda$.

\subsection{Geometrical Interpretation of the Stabilising Transformations $\mathbf{S_{k \sigma}}$} 
\subsubsection{Classification Scheme} 

The stability properties of the FPs of the dynamical systems $S_{k
  \sigma}$ have been investigated in refs.  \cite{Schmelcher97}
exclusively in the context of their relation to the stability
coefficients of the original system $U$. To gain a deeper insight into
the geometrical meaning and the interpretation of the transformations
$S_{k \sigma}$ which turn unstable FP into stable ones, one has to go
beyond the pure consideration of their eigenvalues at the positions of
the FPs.

In the following we develop a geometrical approach allowing us to
classify the FPs which are stabilised by different matrices
${\mathbf{C}}_{k \sigma}$. We will hereby focus on systems with two
degrees of freedom. Expectedly there should, however, be no major
obstacles with respect to the generalisation to arbitrary dimensions.
When dealing with the stability transformations the natural problem
arises: Restricting ourselves to the set of orthogonal ${\mathbf{C}}_{k
  \sigma}$--matrices with exactly one non-vanishing entry $(\pm 1)$ in
each row and column and to the linearised dynamics around a FP, what
can we say about the action of the matrices ${\mathbf{C}}_{k \sigma}$
(see Eqn. (\ref{dynsyss})) on this simple dynamical system?  To
approach this problem consider the following set of equations:
\begin{equation}
\dot{{\mathbf{x}}}=\vec{F}({\mathbf{x}})~~~,~~~~~\vec{F}({\mathbf{x}})
=\vec{f}^{(n)}({\mathbf{x}})-{\mathbf{x}}~~~,~~~~~\vec{F}({\mathbf{x}})=(F_1({\mathbf{x}}),F_2({\mathbf{x}}))^T
\label{contsys}
\end{equation}
which describes a vector field around the FP located at ${\mathbf{x}}_f$,
where $\vec{F}({\mathbf{x}}_f)=\vec{0}$ (the superscript ${}^T$ denotes the transposed).  
In the following sections we generally focus
on the discussion of the system $\vec{F}({\mathbf{x}}_f)$, unless noted differently.
Now we apply the transformation
\begin{equation}
\dot{\mathbf{x}}=\vec{F}_{k \sigma}({\mathbf{x}})={\mathbf{C}}_{k \sigma}\cdot \vec{F}({\mathbf{x}}),
~~~~~\vec{F}_{k \sigma}({\mathbf{x}})=(F_{k \sigma,1}({\mathbf{x}}),F_{k \sigma,2}({\mathbf{x}}))^T
\label{trafocontsys}
\end{equation}
It is important to note that the dynamical system $S_{k \sigma}$ in
Eqn (\ref{dynsyss}) represents a discretisation of the continuous
system (\ref{trafocontsys}).  Multiplication with ${\mathbf{C}}_{k
  \sigma}$ intermingles the $x$-- and $y$--coordinates of
$\vec{F}({\mathbf{x}})$, which in general changes the eigenvalues and
eigenvectors of the corresponding stability matrix. One direction of
the above problem is: Are there any points in the neighbourhood of
${\mathbf{x}}_f$ where this change of the dynamics is controllable?  In
fact let us consider the manifolds $Z_1$, $Z_2$ defined by \cite{Bountis96}

\begin{equation}
Z_i=\{ ~{\mathbf{x}}~| ~F_i({\mathbf{x}})=0~\},~~~ i=1,2
\end{equation}

In the linear neighbourhood of ${\mathbf{x}}_f$ these sets clearly define
straight lines.  In the more general case of a nonlinear system they
are implicitly defined continuous curves in an area of phase space
where the Jacobian of the map is regular.  Their intersection is $Z_1 \cap
Z_2=\{{\mathbf{x}}_f\}$.  With ${\mathbf{C}}_{k \sigma}$ acting on $Z_1$,
$Z_2$, they either stay the same if ${\mathbf{C}}_{k \sigma}$ does not
interchange the coordinates, or they are transferred one onto the
other if ${\mathbf{C}}_{k \sigma}$ does interchange the coordinates.  In
this sense, the manifold $Z=Z_1 \cup Z_2$ is invariant with respect to
application of the set of matrices ${\mathbf{C}}_{k \sigma}$, i.e.
${\mathbf{C}}_{k \sigma}(Z)=Z$ for all $(k \sigma)$.

In the following we derive a classification scheme for the linearised
dynamics around a FP whose validity is however, due to the global
character of our approach, not limited to the linear regime.
In order to elucidate the action of the stabilising transformations
and to distinguish between FPs with different stability
properties, let us introduce two different ways of classifying the
matrices of a two--dimensional dynamical system, each providing its
own insights.  The classifications are such that they reflect certain
geometrical features of the flow around the FP.  These features are
the different invariant sets $Z$ on the one hand and additional
geometrical properties of the matrices which have a particular $Z$ in
common on the other hand.

The first classification introduces classes consisting of matrices
which have the manifold $Z=Z_1\cup Z_2$ in common. They are labelled
${\mathcal{C}}({\phi_{\min},\phi_{\max}})$, where $\phi_{\min}$ and
$\phi_{\max}$ are the azimuthal angles of the manifolds $Z_1$ and
$Z_2$, respectively, being sorted with increasing order.
$\phi_{\min}$ and $\phi_{\max}$ are related to the stability matrix of
the FP in the following way: The linearised dynamics of eqn.
(\ref{contsys}) in the neighbourhood of a FP reads $
\dot{{\mathbf{x}}}={\mathbf{B}}\cdot {\mathbf{x}}$, where
${\mathbf{B}}=(a_{ij})_{1\le i\le 2\atop 1\le j\le 2}$ is the stability
matrix of $\vec{F}({\mathbf{x}})$ at the FP and $ {\mathbf{x}}$ is the
displacement with respect to the FP. It can be shown that
$\phi_{\min}$ and $\phi_{\max}$ are given by
\begin{equation}
\left(\phi_{\min},\phi_{\max}\right)=\left(\min_i\{\phi_i\},\max_i\{\phi_i\}\right)
~~~\mbox{with}~~~ \phi_i=\arctan\left(-\frac{a_{i1}}{a_{i2}}\right)~~~~,~~~~~~i=1,2
\end{equation}
If a stability matrix ${\mathbf{B}}$ belongs to the class
${\mathcal{C}}({\phi_{\min},\phi_{\max}})$ also its products
$\{{\mathbf{C}}_{k\sigma}\cdot {\mathbf{B}}\}$ belong to this class.

For the later on discussion we introduce here three sets of FPs each
of which is an infinite unification of classes
${\mathcal{C}}({\phi_{\min},\phi_{\max}})$ :
\begin{eqnarray}
{\mathcal{C}}^1&=&\{{\mathcal{C}}({\phi_{\min},\phi_{\max}})|~0<\phi_{\min},\phi_{\max}<\pi/2\}\nonumber\\
{\mathcal{C}}^2&=&\{{\mathcal{C}}({\phi_{\min},\phi_{\max}})|~0<\phi_{\min}<\pi/2<\phi_{\max}<\pi\}
\label{defc123}\\ 
{\mathcal{C}}^3&=&\{{\mathcal{C}}({\phi_{\min},\phi_{\max}})|~\pi/2<\phi_{\min},\phi_{\max}<\pi\}\nonumber
\end{eqnarray}
A further classification of the matrices within each class
${\mathcal{C}}({\phi_{\min},\phi_{\max}})$ is needed for a more detailed
identification of the geometrical properties of the flux around a
particular FP.  To this aim, we assign a label $(l \tau)$ to each
stability matrix ${\mathbf{B}}_{l \tau}$ of FPs with the following
meaning: $\tau=\pm1$ gives the sign of $(\det {\mathbf{B}}_{l \tau})$.
To illustrate the meaning of $l$, we write $\vec{F}({\mathbf{x}})=(r\cos
\psi,r\sin\psi)^T$ in polar coordinates and consider the azimuthal
angle $\psi_{\min}$ of $\vec{F}({\mathbf{x}})$ for ${\mathbf{x}}=(\cos
\phi_{\min},\sin\phi_{\min})^T$.  By construction, $\phi_{\min}$ can
take the values $m\pi/2$, $m=0,...,3$. Now we define the index
$l=m+\tau-1 \bmod 4$.  Encircling the FP on a unit circle
${\mathbf{x}}=(\cos \phi,\sin\phi)^T$, the normalised flux
$\vec{F}({\mathbf{x}})/|\vec{F}({\mathbf{x}})|$ describes a circle in the local
coordinate system centred in ${\mathbf{x}}$, too.  $\tau$ gives the
orientation of this circular rotation
of the flux ($\tau=+1$: anticlockwise, $\tau=-1$: clockwise) whereas
$l$ is directly related to the phase of the flux at $\phi_{\min}$.
This naturally introduces a subset ${\mathcal{C}}_{l\tau}(\phi_{\min},\phi_{\max})$
of the class ${\mathcal{C}}(\phi_{\min},\phi_{\max})$: those members of 
${\mathcal{C}}(\phi_{\min},\phi_{\max})$ belong to 
${\mathcal{C}}_{l\tau}(\phi_{\min},\phi_{\max})$ which possess the indices $(l\tau)$,
i.e. the sign $\tau$ for the rotation of the flux and the phase $l$ of the flux at 
$\phi_{\min}$.
We can now allow $\phi_{\min},\phi_{\max}$ to vary while keeping the
indices $(l \tau)$ of the matrices of this set fixed.  For each $(l
\tau)$ we thereby form a class
${\mathcal{A}}_{l\tau}=\{{\mathcal{C}}_{l\tau}({\phi_{\min},\phi_{\max}})|0\le
\phi_{\min},\phi_{\max}\le 2\pi\}$.
For fixed $\phi_{\min},\phi_{\max}$, multiplication by the matrices
${\mathbf{C}}_{k \sigma}$ transfers one complete set
${\mathcal{A}}_{l\tau}$ into another set
${\mathcal{A}}_{l^\prime\tau^\prime}$.  The corresponding transitions
are given in table \ref{fptrans}.  The asterisks in the first three
columns of table \ref{fptrans} indicate the sets ${\mathcal{C}}^i$ {\bf
  and} ${\mathcal{A}}_{l\tau}$ to which stability matrices of a FP of a
two dimensional chaotic system can belong.  This is of relevance when
asking for the possible sets of matrices ${\mathbf{C}}_{k \sigma}$ which
stabilise all FPs of a dynamical system (see below).  As one can read
off table \ref{fptrans}, the matrix ${\mathbf{C}}_{k \sigma}$ necessary
to transfer stability matrices of a class ${\mathcal{A}}_{l\tau}$ into
matrices in another class ${\mathcal{A}}_{l^\prime\tau^\prime}$ is given
by
\begin{equation}
{\mathbf{C}}_{k \sigma}:{\mathcal{A}}_{l\tau}\longrightarrow{\mathcal{A}}_{l^\prime\tau^\prime}
~~\mbox{with}~~l^\prime=k+\sigma\bmod 4~~\mbox{and}~~\tau^\prime=\sigma/\tau
\label{fptranseqn}
\end{equation}
It is important to note that table \ref{fptrans} holds not only for
the unified sets ${\mathcal{A}}_{l\tau}$ but particularly also for the
individual subsets ${\mathcal{C}}_{l\tau}(\phi_{\min},\phi_{\max})$.  It
is a remarkable result that the multiplication table of the matrices
${\mathbf{C}}_{k \sigma}$, i.e.  table \ref{cmult}, considered as a
transition table, has the same entries as the transition table
\ref{fptrans}.  The multiplication law eqn. (\ref{cmulteqn}) becomes a
law for the transition of ${\mathbf{C}}_{k \sigma}$ to
${\mathbf{C}}_{k^{\prime\prime} \sigma^{\prime\prime}}$ via
${\mathbf{C}}_{k^{\prime} \sigma^{\prime}}$ with the corresponding
indices $k^{\prime}=k^{\prime\prime}-\sigma\sigma^{\prime\prime}k$ and
$\sigma^{\prime}=\sigma\sigma^{\prime\prime}$.  This suggests that the
matrices in a class ${\mathcal{A}}_{l\tau}$ are in a way similar to the
matrices ${\mathbf{C}}_{k \sigma}$ with $k=l$ and $\sigma=\tau$ as far
as the rules for multiplication with the matrices ${\mathbf{C}}_{k
  \sigma}$ are concerned.

The class ${\mathcal{C}}_{l\tau}(\phi_{\min},\phi_{\max})$ contains
still an infinite number of (stability) matrices.  However, to gain
relevant information on the stability properties of FPs it suffices,
as we shall see in the following, to know to which of the sets
$\{{\mathcal{C}}^i|i=1...3\}$ and additionally ${\mathcal{A}}_{l\tau}$ the
stability matrix of the FP corresponds.

\subsubsection{Properties of the Angular Functions {$\psi(\phi)$} of the Flux and Examples}

Let us consider an arbitrary but fixed stability matrix
${\mathbf{B}}_{l\tau}\in {\mathcal{C}}(\phi_{\min},\phi_{\max})$,
${\mathbf{B}}_{l\tau}\in {\mathcal{A}}_{l\tau}$, and form the matrices
${\mathbf{C}}_{k \sigma}\cdot{\mathbf{B}}_{l\tau}$,
$k=0,...,3;\sigma=\pm1$.  Each of these matrices is element i.e
representative of a different class
${\mathcal{A}}_{l^\prime\tau^\prime}$ and we therefore have
\begin{equation}
{\mathbf{B}}_{l^\prime \tau^\prime}={\mathbf{C}}_{k \sigma}\cdot{\mathbf{B}}_{l\tau}~~~\mbox{with}~~~ 
l^\prime=k+\sigma l \bmod 4~~\mbox{and}~~\tau^\prime=\sigma/\tau
\label{defblt}
\end{equation}
In the following we will call the set $\{{\mathbf{B}}_{l
  \tau}|l=0,...,3;\tau=\pm1\}$ the family of the matrix
${\mathbf{B}}_{l\tau}$.  A central issue is to analyse which members of
the family of a FP are stable.  Once the information on the stability
of the members of a family becomes available this represents an
important step towards the selective use of eqn. (\ref{dynsyss}), i.e.
the selective detection of FPs.

In the following we study the orientational properties of the flux for
a family of stability matrices with fixed but arbitrary
$(\phi_{\min},\phi_{\max})$.  For the matrices ${\mathbf{B}}_{l
  \tau}=(b^{(l\tau)}_{ij})_{_{1\le i,j\le 2}}\in {\mathcal{A}}_{l\tau}$
we introduce the angular functions
$\psi_{l \tau}(\phi)$ of
the flux $\dot{{\mathbf{x}}}={\mathbf{B}}_{l \tau}\cdot{\mathbf{x}}$ at a point
${\mathbf{x}}=(\cos\phi,\sin\phi)^T$\begin{eqnarray}
\psi_{l \tau}(\phi)&=
&\arctan\frac{({\mathbf{B}}_{l \tau}\cdot{\mathbf{x}})_2}{({\mathbf{B}}_{l \tau}\cdot{\mathbf{x}})_1}+
\pi\left(\frac{3}{2}-\mbox{sign}({\mathbf{B}}_{l \tau}\cdot{\mathbf{x}})_2\right)\label{defpsi}\\
&=&\arctan\left(\frac{b^{(l\tau)}_{21}\cos\phi+b^{(l\tau)}_{22}\sin\phi}{b^{(l\tau)}_{11}\cos\phi+b^{(l\tau)}_{12}\sin\phi}\right)
+\pi\left(\frac{3}{2}-\mbox{sign}\left(b^{(l\tau)}_{21}\cos\phi+b^{(l\tau)}_{22}\sin\phi\right)\right)\nonumber
\end{eqnarray}
$\psi_{l \tau}(\phi)$ is the azimuthal angle of the flux (given by
$\vec{F}_{l \tau}({\mathbf{x}})$) with respect to a local polar coordinate
system centred at the displacement ${\mathbf{x}}=(\cos \phi, \sin
\phi)^T$. This is illustrated in Fig. \ref{coordsys}.  Let us derive relevant properties
of the angular flux functions $\psi_{l \tau}(\phi)$.  $\psi_{l
  \tau}(\phi)$ is a continuous function of $\phi$ with $\psi_{l
  \tau}(\phi)\in [0,2\pi]$.  Due to symmetry reasons it is sufficient
to consider the range $0<\phi<\pi$. Furthermore $\psi_{l \tau}(\phi)$
is defined $\bmod~2\pi$ and $\psi_{l \tau}(0)-\psi_{l
  \tau}(\pi)=\pi$.  Its derivative reads
\begin{equation}
\psi^\prime_{l \tau}(\phi)=\frac{\det({\mathbf{B}}_{l \tau})}{L_{l \tau}^2(\phi)}\label{derivpsi}
\end{equation}
with
\begin{equation}L_{l \tau}(\phi)=|\dot{{\mathbf{x}}}|=
\sqrt{(b^{(l\tau)}_{11}\cos\phi+b^{(l\tau)}_{12}\sin\phi)^2+(b^{(l\tau)}_{21}\cos\phi+b^{(l\tau)}_{22}\sin\phi)^2}
\label{defL}
\end{equation}
being the ``velocity'' of the flux in ${\mathbf{x}}$.  Since
multiplication with any ${\mathbf{C}}_{k \sigma}$ affects complete rows
of ${\mathbf{B}}_{l \tau}$ by interchanging them or inverting their
signs, $L^2_{l \tau}(\phi)$ is invariant, i.e. the same for any of the
resulting matrices ${\mathbf{B}}_{l \tau}$.  Two functions
$\psi_{l\tau}(\phi)$ and $\psi_{l^\prime\tau}(\phi)$ therefore differ
only by a shift: $\psi_{l
  \tau}(\phi)-\psi_{l^\prime\tau}(\phi)=(l-l^\prime \bmod 4)\cdot
\frac{\pi}{2}$.  $l$ can be interpreted (see also subsection 2.2.1) as
the phase of the linearised flux around the FP:
 \begin{equation}
\psi_{l \tau}(\phi_{\min})=(l+1-\tau)\cdot \frac{\pi}{2}
\end{equation}
Accordingly we have $\det({\mathbf{C}}_{k
  \sigma}\cdot{\mathbf{B}})=\det({\mathbf{C}}_{k
  \sigma})\cdot\det({\mathbf{B}})$, with $\det({\mathbf{C}}_{k +})=+1$ and
$\det({\mathbf{C}}_{k -})=-1$ and therefore
\begin{equation}
\psi^\prime_{l \tau}(\phi)=-\psi^\prime_{l -\tau}(\phi)
\end{equation}
which yields
\begin{equation}
\psi_{l \tau}(\phi)=2\cdot \psi_{l \tau}(\phi_{\min})-\psi_{l -\tau}(\phi)
\end{equation}
$\tau=\pm 1$ indicates the sign of $\det({\mathbf{B}}_{l \tau})$ and
determines whether $\psi_{l \tau}(\phi)$ is rising or falling,
according to eqn. (\ref{derivpsi}).

In the remaining part of this subsection we provide generic examples
and illustrations of the linearised dynamics around a FP and the
corresponding angular functions of the flux for the different members
of a family $\{{\mathbf{B}}_{l\tau}\}$.

The stability matrices ${\mathbf{B}}_{l \tau}$ in Figs. \ref{ppc1p}a)
and Fig. \ref{ppc1n}a) are members of a family of the matrix
${\mathbf{M}}_1=\left({1\atop-10} {-4 \atop 5}\right)$, corresponding to
the class ${\mathcal{C}}(0.24,1.10)$.  Figs. \ref{ppc2p}a) and
\ref{ppc2n}a) are obtained in the same way, showing the family of
${\mathbf{M}}_2=\left({1\atop10}{-4\atop5}\right)$, which is in the
class ${\mathcal{C}}(0.24,2.03)$.  Each subfigure shows the linear
neighbourhood of the corresponding FP. $( x, y)$ are the coordinates
of the displacement of a trajectory with respect to the FP. The
manifolds $Z_1$, $Z_2$ are displayed as long dashed lines with the
direction of the flux on the $Z_1$, $Z_2$--lines being indicated with
open arrows.  The thick lines with filled arrows show the position and
stability properties of the eigenvectors (For saddle points,
$\vec{v}_e$, $\vec{v}_c$ are indicated by outward and inward looking
arrows corresponding to the expanding and contracting manifolds,
respectively.  For sinks and sources the corresponding eigenvectors
are labelled $\vec{v}_{e,1}$, $\vec{v}_{e,2}$ and $\vec{v}_{c,1}$,
$\vec{v}_{c,2}$, respectively).  Additionally, some trajectories have
been included to visualise the direction of the flow around the FP.
In Figs. \ref{ppc1p}b) -- \ref{ppc2n}b) the corresponding angular
functions $\psi_{l\tau}(\phi)$ are plotted. It is important to note
that these particular functions -- as well as the phase portraits
given in the corresponding figures -- are merely examples to
demonstrate the qualitative variation of the dynamics generated by
members of a family of stability matrices. There exist in general
other families in the same class
${\mathcal{C}}(\phi_{\min},\phi_{\max})\cap{\mathcal{A}}_{l\tau}$ whose
phase portraits and functions $\psi_{l\tau}(\phi)$ look different from
those displayed in the figures.  However, the functions
$\psi_{l\tau}(\phi)$ of all these matrices in
${\mathcal{C}}(\phi_{\min},\phi_{\max})\cap{\mathcal{A}}_{l\tau}$ have the
values of $\psi_{l\tau}(\phi)$ in $\phi=\phi_{\min}$ and
$\phi=\phi_{\max}$ and the sign of $\psi^\prime(\phi)_{l\tau}$ in
common.  The actual $\psi_{l\tau}(\phi)$ can vary in between according
to the definition (\ref{defpsi}).  Nevertheless the above information
about the stability matrices is in fact sufficient to determine their
stability properties.

Now, looking at the examples in Figs. \ref{ppc1p}a) -- 
\ref{ppc2n}a) except the cases of spiral points ${\mathbf{B}}_{1+}$ 
and ${\mathbf{B}}_{3+}$, we are already in the position to
suggest a criterion for the stability and the approximate position of
the eigenvectors of the FP.  Consider the direction of the flux in
each of the sectors which are determined by the manifolds $Z_1$, $Z_2$
and the coordinate axes.  The manifolds $Z_1$, $Z_2$ are by definition
the lines for which the first, respectively second component of the
flux vanishes.  Since the flux $\vec{F}({\mathbf{x}})$ is a continuous
vector function of the angle $\phi$ there have to be certain values
$\phi_e$ for which the flux $\vec{F}({\mathbf{x}})$ is collinear with the
position vector ${\mathbf{x}}=(\cos \phi, \sin \phi)^T$, i.e.
$\vec{F}({\mathbf{x}})=\pm c\cdot {\mathbf{x}}$, $c>0$.  These angles $\phi_e$
obviously are the polar angles of the eigenvectors of the
corresponding stability matrix.  The intervals where these angles
$\phi_e$ are located are bounded by $Z_1$ and $Z_2$.  The manifolds
$Z_1$ and $Z_2$ represent the boundaries of sectors in which the angle
of the flux varies by $\pi/2$.  For reasons of continuity
of $\psi_{l\tau}(\phi)$ ( see Figs. \ref{ppc1p}a)-- \ref{ppc2n}a)) one
or even two eigenvectors are located within these sectors, which are
shaded grey in the figures.  In specific cases parts of the sectors
defined by $Z_1$ and $Z_2$ are excluded for the localisation of the
eigenvectors since collinearity of ${\mathbf{x}}$ and $\vec{F}({\mathbf{x}})$ is
not possible. In these cases the outer coordinate system $(x,y)$
represents the boundary of the sectors for the occurrence of
collinearity.

The fact that $\phi_e$ is the polar angle of an eigenvector of the
stability matrix ${\mathbf{A}}$ is equivalent to
\begin{equation}
\psi_{l \tau}(\phi_e)=\left\{
\matrix{
\phi_e&~:~~\mbox{Re}(\lambda)>0,&\mbox{unstable eigenvector}\cr
\phi_e+\pi&~:~~\mbox{Re}(\lambda)<0,&\mbox{stable eigenvector}
}
\right.
\end{equation}
This means that a crossing of $\psi_{l \tau}(\phi)$ with
\begin{equation}
\chi_n(\phi)=\phi+n\cdot\pi~~,~~~n=0,1
\label{defpsin}
\end{equation}
indicates an unstable or stable eigenvector for the corresponding
value of $\phi$, respectively.

\subsubsection{Stability Properties of the Classes and the Alternative Sets of Stabilisation Transformations}

We will show in the following that for fixed $\phi_{\min}$ and
$\phi_{\max}$ two complete classes ${\mathcal{A}}_{l^\prime\tau^\prime}$
and ${\mathcal{A}}_{l^{\prime\prime},\tau^{\prime\prime}}$ are related
to stable FPs, i.e. the matrices in these classes have eigenvalues
$\lambda_1$, $\lambda_2$ for which $\mbox{Re}(\lambda_{1}),\mbox{Re}(\lambda_{2})<0$ holds.
To these two classes correspond two matrices ${\mathbf{C}}_{k^\prime
  \sigma^\prime}$ and ${\mathbf{C}}_{k^{\prime\prime}
  \sigma^{\prime\prime}}$ which transfer an original stability matrix
in a class ${\mathcal{A}}_{l\tau}$ into the desired stable matrices in
${\mathcal{A}}_{l^\prime \tau^\prime}$ and
${\mathcal{A}}_{l^{\prime\prime}\tau^{\prime\prime}}$.  This corresponds
to a transformation of the original FP into the desired stable ones in
the transformed dynamical systems $S_{k^\prime \sigma^\prime}$ and
$S_{k^{\prime\prime} \sigma^{\prime\prime}}$.  The two matrices
${\mathbf{C}}_{k^\prime \sigma^\prime}$ and
${\mathbf{C}}_{k^{\prime\prime} \sigma^{\prime\prime}}$ which accomplish
this transformation can be read off the transition table \ref{fptrans}
immediately.  It will in particular become evident that a minimal set
of three ${\mathbf{C}}_{k \sigma}$ matrices is sufficient to stabilise
any FP of a two dimensional fully chaotic system.  What is more, this
classification proves to be useful not only for saddle points but also
for systems with repellors and spiral points, which can be transformed
to stable FPs (sinks) via a certain matrix ${\mathbf{C}}_{k \sigma}$
(see below).

We now discuss the properties of all possible stability matrices
according to their $\psi_{l\tau}(\phi)$--diagrams. To do this it is
sufficient to consider the assignment of the matrices to the classes
${\mathcal{C}}^1$, ${\mathcal{C}}^2$ and ${\mathcal{C}}^3$ as introduced in
(\ref{defc123}) and to the classes ${\mathcal{A}}_{l\tau}$. In the
following we concentrate on the classes ${\mathcal{C}}^1$ and
${\mathcal{C}}^2$ since the argumentation for the class ${\mathcal{C}}^3$
is analogous to that for ${\mathcal{C}}^1$. For the same reason we
restricted the examples in Figs. \ref{ppc1p}--\ref{ppc2n} to the
classes ${\mathcal{C}}^1$ and ${\mathcal{C}}^2$.  In Figs.
\ref{ppc1p}c)--\ref{ppc2n}c) we show the areas in the $\psi_{l
  \tau}(\phi)$--diagrams where a crossing of $\psi_{l \tau}(\phi)$ and
$\chi_0(\phi)$ or $\chi_1(\phi)$ may occur as gray shaded boxes.  They
are derived by simple application of continuity arguments concerning
$\psi_{l \tau}(\phi)$.  The labels $l\tau$ in the boxes are the labels
of the corresponding class ${\mathcal{A}}_{l\tau}$ of matrices whose
real eigenvectors have azimuthal angles in this range of $\phi$.  Two
labels given in brackets indicate the possibility of either two real
eigenvalues with eigenvectors in this range (sink or source) or
complex eigenvalues without real eigenvectors of the corresponding
matrix (spiral points).  These two possibilities cannot be
distinguished within our classification scheme of matrices.  However,
this fact does not affect the final selection of the minimal
stabilising set of matrices ${\mathbf{C}}_{k\sigma}$.

We begin our discussion with stability matrices ${\mathbf{B}}$ with
negative determinants:
\begin{itemize}
\item Matrices ${\mathbf{B}}\in\{{\mathcal{A}}_{l-}|l=0...3\}$ are
  stability matrices of saddle points.  The stability properties of
  these FPs are easy to determine from the corresponding $\psi_{l
    \tau}(\phi)$-- diagrams in Figs. \ref{ppc1n}b), \ref{ppc2n}b).  It
  is obvious from the monotonicity and continuity of the $\psi_{l
    \tau}(\phi)$--curve that they intersect the ``unstable'' and the
  ``stable'' lines $\chi_0(\phi)$, $\chi_1(\phi)$, respectively,
  exactly once.  The sectors where the corresponding eigenvectors are
  localised are shaded grey in the corresponding phase diagrams Figs.
  \ref{ppc1n}c), \ref{ppc2n}c).
\end{itemize}

FPs with positive Jacobian ${\mathbf{B}}$ are a bit more subtle matter.
They belong to the classes $\{{\mathcal{A}}_{l +}|l=0...3\}$ and include
sinks and sources as well as spiral points for which the real part of
the stability eigenvalues $\mbox{Re}(\lambda_1)$
$\mbox{Re}(\lambda_2)$ possess the same signs.  Spiral FPs possess
stability matrices ${\mathbf{B}}$ with eigenvalues
$\mbox{Im}(\lambda_1),\mbox{Im}(\lambda_2)\not=0$.  This implies
$\det({\mathbf{B}})>0$.  As pointed out earlier the classification of
the stability matrices with ${\mathcal{C}}(\phi_{\min},\phi_{\max})$ and
${\mathcal{A}}_{l\tau}$ does neither specify the matrix nor its
eigenvalues completely.  In particular for stability matrices whose
family contains spiral points an additional criterion is needed to
analyse their stability since the corresponding $\psi_{l
  \tau}(\phi)$--functions do not cross the lines $\chi_0(\phi)$ or
$\chi_1(\phi)$.  For this analysis we suggest the following criterion
(whose completeness we could not prove yet).
 
Consider the angles $\phi_t$, $t=1,2$, for which $\psi^\prime_{l
  \tau}(\phi_t)=1$, i.e.  $\psi_{l \tau}(\phi)$ is tangential to
$\chi_n(\phi)$, and define the distance to both lines $\chi_{n}(\phi)$
\begin{equation}
d_n=\min_{t=1,2}\left|\psi_{l \tau}(\phi_t)-\phi_t-n\pi\right|,~~~n=0,1
\label{dn}
\end{equation}
Now the stability of the spiral point is suggested by
\begin{eqnarray}
d_0<d_1&\Longrightarrow&\mbox{Re}(\lambda_e)>0,~~~~~\mbox{unstable spiral point}\\
d_0>d_1&\Longrightarrow&\mbox{Re}(\lambda_e)<0,~~~~~\mbox{stable spiral point}
\end{eqnarray}
This implies that for spiral FPs the line $\chi_n(\phi)$ which is
closest to $\psi_{l \tau}(\phi_t)$ determines the stability and this
can be seen as a generalisation of the criterion of the crossing with
$\chi_n(\phi)$ in the case of real eigenvalues.

It is an interesting property that the stability of matrices
${\mathbf{B}}_{l +}$ of a family
$\{{\mathbf{B}}_{l\tau}|l=0...3,\tau=\pm\}$ changes with $l\to l+2\bmod
4$, i.e. if ${\mathbf{B}}_{1 +}$ is stable, then ${\mathbf{B}}_{3 +}$ is
unstable, if ${\mathbf{B}}_{2 +}$ is stable, then ${\mathbf{B}}_{4 +}$ is
unstable and vice versa.  If the eigenvalues are real the eigenvectors
are not affected by the shift (only their stability changes).  This
property is obvious taking into account that the shift in $l$
corresponds to a shift of $\psi_{l\tau}(\phi)$ by $\pi$, with
$\pi$ being also the shift between the lines $\chi_0(\phi)$,
$\chi_1(\phi)$.  Physically, the shift by $\pi$ is for any stability
matrix equivalent to a time reversal (which corresponds to a
reversal of the flux $\vec{F}({\mathbf{x}})\to -\vec{F}({\mathbf{x}})$.

We now come back to the discussion of the stability properties of
matrices ${\mathbf{B}}$ possessing positive determinant:

\begin{itemize}
\item Matrices ${\mathbf{B}}\in\{{\mathcal{A}}_{l +}|l=0...3\}$ describe
  sinks and sources and spiral points.  
  We first address the sinks and sources and second the spiral points.
  The corresponding curve $\psi_{l\tau}(\phi)$ crosses one of the lines 
  $\chi_0(\phi)$ (sink)
  or $\chi_1(\phi)$ (source) twice.  It is obvious but nevertheless 
  important to note that
  there cannot be more than two crossings,  
  which can also formally be seen from eq.  (\ref{derivpsi}).  Having
  three or more cuts with $\chi_n(\phi)$, $n=0,1$ implies that
  $\psi_{l\tau}(\phi)$ has at least two turning points in any interval
  $[\phi_d,\phi_u]$ with $\phi_d-\phi_u=\pi/2$.  This means that
  $L_{l \tau}^2(\phi)$ has two or more extrema in $[\phi_d,\phi_u]$.
  Differentiating $L^2_{l \tau}(\phi)$ (see eqn. (\ref{defL}))
yields (for the sake of simplicity we omit the superscripts $(l\tau)$ in
the entries $\left(b_{ij}^{(l\tau)}\right)_{1\le i,j\le 2}$ of ${\mathbf{B}}$)
\begin{eqnarray}
\frac{\mbox{d}~L_{l \tau}^2(\phi)}{\mbox{d}~\phi}&=&
-(b_{21}^2+b_{11}^2)2\cos\phi\sin\phi+(b_{22}^2+b_{12}^2)2\sin\phi\cos\phi+\nonumber\\
&&2(\cos^2\phi-\sin^2\phi)(b_{21}b_{22}+b_{11}b_{12})\nonumber\\
&=&(b_{22}^2+b_{12}^2-b_{21}^2-b_{11}^2)\sin(2\phi)+\nonumber\\
&&2(\cos^2\phi-\sin^2\phi)(b_{21}b_{22}+b_{11}b_{12})\cos(2\phi)
\end{eqnarray}
\begin{equation}\frac{\mbox{d}~L_{l \tau}^2(\phi)}{\mbox{d}~\phi}=0~~~~\Longleftrightarrow~~~~
\tan(2\phi)=\frac{b_{21}b_{22}+b_{11}b_{12}}{b_{22}^2+b_{12}^2-b_{21}^2-b_{11}^2}
\end{equation}
which has exactly one solution for $\phi$ in any interval
$[\phi_d,\phi_u]$ with $\phi_d-\phi_u=\pi/2$, i.e. there is only one
turning point in this interval.

\begin{itemize}
\item \underline{The class ${\mathcal{C}}^1$:}
As can be read off
directly from the diagrams Figs. \ref{ppc1p}b) and c), matrices in
${\mathcal{C}}^1\cap{\mathcal{A}}_{2+}$ and
${\mathcal{C}}^1\cap{\mathcal{A}}_{2-}$ are sinks and sources,
respectively. One of the eigenvectors is in the interval
$[\phi_{\min},\phi_{\max}]$, the other one is located in $[\pi/2,\pi]$.
Matrices of the class ${\mathcal{C}}^1\cap{\mathcal{A}}_{1+}$ and
${\mathcal{C}}^1\cap{\mathcal{A}}_{3+}$ are either sinks and sources or
spiral points. 
For real eigenvalues matrices of ${\mathcal{C}}^1\cap{\mathcal{A}}_{1+}$
are sinks whereas matrices in
${\mathcal{C}}^1\cap{\mathcal{A}}_{3 +}$ are sources.  The orientation 
of the eigenvectors of
both stability matrices are within the interval $[\phi_{\max},\pi/2]$.

Considering the latter case of spiral points, we can at least say that
within one family $\{{\mathbf{B}}_{l\tau}|l=0...3,\tau=\pm1\}$ of
matrices either ${\mathbf{B}}_{1+}$ is a stable spiral and
${\mathbf{B}}_{3+}$ is an unstable spiral point or vice versa.

\item \underline{The class ${\mathcal{C}}^2$:}
Here more cases are
possible. Looking at the matrices $\{{\mathbf{B}}_{l+}|l=0...3\}$ of the
family $\{{\mathbf{B}}_{l\tau}|l=0...3,\tau=\pm\}$ there can be either
\begin{itemize}
\item no spiral point, but two sinks and two sources (analogous to
  ${\mathcal{C}}^1$)
\item two spiral points (one stable, one unstable), one sink and one
  source (analogous to in ${\mathcal{C}}^1$)
\item four spiral points (two stable, two unstable)
\end{itemize}
Which of these cases occur is a question of the variation of $\psi_{l
  \tau}(\phi)-\phi$ and of the phase $\psi_{l
  \tau}(\phi_{\min})$.  If real eigenvalues occur, the angles of the
corresponding eigenvectors are within $[\pi/2,\phi_{\max}]$ for
matrices in ${\mathcal{A}}_{1+}$ and ${\mathcal{A}}_{3+}$. The position
eigenvectors of stability matrices matrices in ${\mathcal{A}}_{1+}$ and
${\mathcal{A}}_{3+}$ is not determined within this classification
scheme.
\end{itemize}
\end{itemize}

To determine the minimal sets ${\mathcal{S}}_i$ of matrices
${\mathbf{C}}_{k\sigma}$ necessary for stabilisation of {\bf all} FPs of a two
dimensional chaotic dynamical system let us first consider the classes
${\mathcal{A}}_{l\tau}\cap{\mathcal{C}}^i$ whose representatives correspond to 
stable FPs.
We herefore form pairs
$(l\tau,l^\prime\tau^\prime)$ abbreviating the two classes
${\mathcal{A}}_{l\tau}$ and ${\mathcal{A}}_{l^\prime\tau^\prime}$ 
for any ${\mathcal{C}}^i$.
\begin{eqnarray}
{\mathcal{C}}^1&:&(1+,2+) ~~\mbox{or}~~ (2+,3+)\nonumber\\
{\mathcal{C}}^2&:&(1+,2+) ~~\mbox{or}~~ (2+,3+)~~\mbox{or}~~ (3+,0+)\label{ssaddle}
\\
{\mathcal{C}}^3&:&(1+,2+) ~~\mbox{or}~~ (2+,3+)\nonumber
\end{eqnarray}
When looking for the minimal set of matrices ${\mathbf{C}}_{k\sigma}$
necessary for stabilisation one has to take into account that only certain
kinds of FPs can occur in the system $\dot{{\mathbf{x}}}=\vec{F}({\mathbf{x}})$,
eqn. (\ref{contsys}), derived from the original dynamical system
${\mathbf{x}}_{i+1}=\vec{f}({\mathbf{x}}_i)$. In two dimensions $\vec{f}({\mathbf{x}})$
can have saddle points only and the FPs of $\vec{F}({\mathbf{x}})$
are therefore either saddle points too or sinks.

Let us discuss the set ${\mathcal{S}}_{saddle}$ of matrices
${\mathbf{C}}_{k\sigma}$ which stabilise saddle points first. Since the
determinant of the stability matrix is negative for a saddle point and
positive for any stable sink or spiral point, the corresponding stabilising matrix
${\mathbf{C}}_{k\sigma}$ has the form ${\mathbf{C}}_{k-}$. Since the saddle
points of all classes ${\mathcal{C}}^1$, ${\mathcal{C}}^2$ and
${\mathcal{C}}^3$ have to be stabilised by ${\mathcal{S}}$ we have to
determine ${\mathcal{S}}$ such that any class ${\mathcal{A}}_{l-}$ of
original matrices is transferred into at least one element in each of the pairs
$(1+,2+) ~~\mbox{or}~~ (2+,3+)~~\mbox{or}~~ (3+,0+)$,
which is the union of the pairs of all three classes
${\mathcal{C}}^1$, ${\mathcal{C}}^2$ and ${\mathcal{C}}^3$ in eqn.
(\ref{ssaddle}).  The transition table \ref{fptrans} shows that
there are just two possibilities of minimal sets:
${\mathcal{S}}_{saddle}=\{{\mathbf{C}}_{0-},{\mathbf{C}}_{2-}\}$ or
$\{{\mathbf{C}}_{1-},{\mathbf{C}}_{3-}\}$.  Each of these sets
has to be combined with sets
${\mathcal{S}}_{sink}$ that stabilise the sinks of
$\dot{{\mathbf{x}}}=\vec{F}({\mathbf{x}})$ to yield a possible set
${\mathcal{S}}$. 
Since the latter are already stable, the identity transformation 
${\mathcal{S}}_{sink}=\{{\mathbf{C}}_{0+}\}$ is sufficient. Indeed, it 
is easy to see that no other
${\mathbf{C}}_{k\sigma}$ is able to achieve the same: Sinks can occur in
both ${\mathcal{C}}^1$ and ${\mathcal{C}}^3$ for the classes
${\mathcal{A}}_{1+}$ and ${\mathcal{A}}_{2+}$ and in ${\mathcal{C}}^2$ for the
classes ${\mathcal{A}}_{2+}$ and ${\mathcal{A}}_{3+}$.  We therefore can
list these different classes of sinks and sets ${\mathcal{S}}_{sink}$ of
matrices ${\mathbf{C}}_{k\sigma}$ necessary to stabilise them as
follows:
\begin{equation}
\matrix{
\mbox{original classes}&
\mbox{stable classes}&\mbox{sets of stabilising~}~{\mathbf{C}}_{k\sigma}\cr
{\mathcal{C}}^1\cap{\mathcal{A}}_{1+}:&(1+,2+)&\{{\mathbf{C}}_{0+}\}~\mbox{or}~\{{\mathbf{C}}_{1+}\}\cr
{\mathcal{C}}^1\cap{\mathcal{A}}_{2+}:&(1+,2+)~\mbox{or}~(2+,3+)
&\{{\mathbf{C}}_{0+}\}~\mbox{or}~\{{\mathbf{C}}_{1+},{\mathbf{C}}_{3+}\}\cr
{\mathcal{C}}^2\cap{\mathcal{A}}_{2+}:&(1+,2+)~\mbox{or}~(2+,3+)
&\{{\mathbf{C}}_{0+}\}~\mbox{or}~\{{\mathbf{C}}_{1+},{\mathbf{C}}_{3+}\}\cr
{\mathcal{C}}^2\cap{\mathcal{A}}_{3+}:&(2+,3+)~\mbox{or}~(0+,3+)
&\{{\mathbf{C}}_{0+}\}~\mbox{or}~\{{\mathbf{C}}_{1+},{\mathbf{C}}_{3+}\}\cr
{\mathcal{C}}^3\cap{\mathcal{A}}_{1+}:&(1+,2+)&\{{\mathbf{C}}_{0+}\}~\mbox{or}~\{{\mathbf{C}}_{1+}\}\cr
{\mathcal{C}}^3\cap{\mathcal{A}}_{2+}:&(1+,2+)~\mbox{or}~(2+,3+)
&\{{\mathbf{C}}_{0+}\}~\mbox{or}~\{{\mathbf{C}}_{1+},{\mathbf{C}}_{3+}\}\cr
}
\end{equation}
So the smallest set that stabilises any sink independent of the
classes ${\mathcal{C}}^1$, ${\mathcal{C}}^2$ and ${\mathcal{C}}^3$ is
${\mathcal{S}}_{sink}=\{{\mathbf{C}}_{0+}\}$.  If we now form the
unions ${\mathcal{S}}_{saddle}\cup{\mathcal{S}}_{sink}$ to get a set of
matrices ${\mathbf{C}}_{k\sigma}$ that stabilise all FPs of a
two dimensional fully chaotic dynamical system we end up with two 'global'
minimal sets:
\begin{equation}
{\mathcal{S}}_1=\{{\mathbf{C}}_{0+},{\mathbf{C}}_{0-},{\mathbf{C}}_{2-}\}~~~\mbox{and}~~~
{\mathcal{S}}_2=\{{\mathbf{C}}_{0+},{\mathbf{C}}_{1-},{\mathbf{C}}_{3-}\}
\end{equation}
There are other sets which also do the job, but they
contain at least four matrices and are therefore not minimal.
In previous numerical investigations it was observed\cite{Schmelcher97,Diakonos98} that the transformation
belonging to the matrix ${\mathbf{C}}_{0+}$ yields a particularly large number of
stabilized fixed points. From the above discussion this becomes now understandable
since ${\mathbf{C}}_{0+}$ is responsible for the stabilization of the sinks in equation 
(\ref{trafocontsys}) which occur in many different classes.

Let us remark that the above discussion includes also sinks of the original 
system $\vec{f}$ since they become sinks of $\vec{F}$ and can therefore
be treated in the same way, i.e. are 'stabilised' using the same minimal sets ${\mathcal{S}}_1$
or ${\mathcal{S}}_2$.

The concept of characterising FPs by sets of manifolds which are 
invariant with respect to the 
operations ${\mathbf{C}}_{k\sigma}$ can be extended in a natural way. 
One can consider manifolds
which are composed by not
only two subsets $Z_1$ and $Z_2$ as above, but of four subsets $Z_1...Z_4$. 
These subsets are defined implicitly by demanding a given ratio of 
the two components of the flux. 
They correspond for the linear regime to four angles 
$\{\phi_1,\phi_2,\phi_3,\phi_4\}$, which are a generalisation of the parameters 
$\{\phi_{\min},\phi_{\max}\}$ in the discussion above. 
This larger set of parameters leads to a 
finer partition of the space of all $2\times 2$ matrices, which in turn allows 
a complete assignment 
of stability properties to the different matrices. 
The definition of the family of a given matrix is analogous to the corresponding definition
(\ref{defblt}) and reflects the action of the group of matrices $\{{\mathbf{C}_{k\sigma}}\}$.
However, finding the proper partition is not straightforward and is left to 
a future investigation.

\section{Applications}

In this section we apply the SD method whose theoretical background has been discussed 
so far. Furthermore we will discuss several improvements of its original 
algorithmic implementation to locate the UPOs \cite{Schmelcher97,Diakonos98}
With respect to the detection of the orbits our aim is twofold. First we are interested in 
complete sets of UPOs for higher periods of the above maps and in analysing them with respect
to the distributions of the corresponding Lyapunov exponents. Second we
want to demonstrate the suitability of our method to detect the least unstable orbits
up to high periods. The latter is an extension of the work given in ref. \cite{Diakonos98}.
We remark that very recently an efficient algorithm for detecting UPOs in chaotic systems based on
a combination of the SD method and a Newton-Raphson like approach has been developed
and successfully applied \cite{Lai99}.

\subsection{Finding the Fixed Points}

Concerning the efficient algorithmic implementations of the SD method we face two main problems:
\begin{itemize}
\item[1)] The completeness of the detected set of UPOs which is of conceptual character.
\item[2)] Separating closely neighboured UPOs which is an issue only for our particular
implementation of the SD method, i.e. characteristic for our numerical approach.
\end{itemize}
Let us first address the completeness problem. 
Of course there is no exact proof of completeness for the detected UPOs within
the SD method. However, a properly chosen sequence of sets of initial conditions which cover
the phase space of the dynamical system as neat as possible can significantly lower the 
probability of missing any UPOs.
Due to the
presence of length scales which differ by many orders of magnitude
induced by the fractal structure of the corresponding strange
attractor we proceed here as follows.

We introduce a set of grids $G_i$, i=1,2,...  of starting points which
are cumulative in the way that the points of $G_{i}$ fill gaps on the attractor that
are larger
than a given size in the union of the preceding grids $G_{1}\cup
G_{2}\cup...\cup G_{i-1}$.  In the particular case of the Ikeda map (see below) we
generate a sequence of six grids $G_1$,...,$G_6$ with $G_1$ containing
approximately 4500 points, while $G_i$, $i=2,...,6$ contain
approximately 1500 points each.
The starting points of each set $G_i$ are propagated with the
transformed maps $S_{0+}$, $S_{0-}$, and $S_{2-}$, i.e. applying the
matrices ${\mathbf{C}}_{0+}$, ${\mathbf{C}}_{0-}$ and ${\mathbf{C}}_{2-}$
according to (\ref{dynsyss}).  The propagation of a particular
trajectory is stopped at the point ${\mathbf{x}}$ if
$\delta=|f^{(p)}({\mathbf{x}})-{\mathbf{x}}|<\epsilon$, where $\epsilon$ is the
desired resolution of the FPs.  
For periodic orbits of the Ikeda map with
$p=14$ and $15$, $\epsilon<10^{-10}$ turned out to be sufficient to
resolve the different cycles.  
Propagating the starting points of the grids $G_i$ with
$S_{0+}$, $S_{0-}$ and $S_{2-}$ yields the sets $N_{i,0+}$, $N_{i,0-}$
and $N_{i,2-}$, $i=1,...,6$, respectively, of points which to good accuracy approximate
the FPs of the Ikeda map.
The set $N_i = N_{i,0+}\cup N_{i,0-}\cup N_{i,2-}$ contains the FPs of
the map found by propagating the points of $G_i$.  Then we consider the
number $n_{i,k \sigma}$ of FPs that appear in a particular $N_{i,k
  \sigma}$ (and therefore in $N_i$) with $( k \sigma) = (0+)$, $(0-)$,
$(2-)$, but which are not contained in any other $N_j$, $j<i$.  If
$n_{i-1,k \sigma}=n_{i,k \sigma}=0$, i.e. no additional FPs have been
found when propagating two subsequent grids $G_{i-1}$, $G_{i}$, the
set of FPs of the transformed system $S_{k \sigma}$ is considered to
be complete and the procedure of constructing subsequent grids is stopped.

The second improvement concerns the separation of neighbouring UPOs.  
Using however an appropriately defined distance $d_{xy}$
between two orbits of period $p$, $\vec{x_i}=(x_{1i},x_{2i})^T$,
$\vec{y_i}=(y_{1i},y_{2i})^T$, $i=1,\dots,p$:
\begin{equation}
d_{xy}=\min_{k=0,\dots\atop \ldots,p-1}\sum_{1\le i\le p\atop 1\le j\le 2}(x_{ji}-y_{j(i+k\bmod p)})^2
\label{dist}
\end{equation}

In the following we consider a set of FPs that belong to
different periodic orbits.  The FPs obtained in the procedure of
propagating the transformed systems $S_{0+}$, $S_{0-}$ and $S_{2-}$ as
described above provide an example for such a set.  By forming all
possible pairs of any two FPs of this set and looking at the resulting
distribution of values of $d_{xy}$ (e.g. by plotting all $d_{xy}$ in
a logarithmic scale) one can distinguish three subsets, separated by
gaps differing by several orders of magnitude (Fig. \ref{dxyplot}).
The set with the largest values of $d_{xy}$ contains all pairs formed with
different orbits, whereas the second and third largest set is composed of
pairs of identical orbits.
A possible explanation for the appearance of the gap between the
second and the third zone is the following:
A trajectory 
of the transformed map approaches the FP on a curve which in the local but
still nonlinear neighbourhood
is close to the least stable of both stable manifolds of the FP. 
Two trajectories $\{{\mathbf{y}}_i\}$ and $\{{\mathbf{z}}_i\}$
can therefore evolve towards a FP $\mathbf{x}_o$ along the 
same or opposite directions. The same holds for all other FPs 
of the orbit
$\vec{f}^{(r)}({\mathbf{x}}_o)$ with the trajectories $\{\vec{f}^{(r)}({\mathbf{y}}_i)\}$ and
$\{\vec{f}^{(r)}({\mathbf{z}}_i)\}$, $r=1,...,p-1$.
In the case of an antiparallel approach to the FPs of a given cycle the contributions to 
$d_{xy}$ are much larger compared to the case of a parallel approach,
which finally yields the gap between the second (parallel approach ) and third
(antiparallel approach) region in Fig. \ref{dxyplot}.
If the first and the second zone happen to merge, the accuracy of the FPs is not
sufficient to distinguish different orbits and has to
be refined by further propagation with the corresponding transformed
maps.
What is more, this distinction can be used to derive an estimate for
the absolute accuracy of the FPs derived by propagating the
appropriately transformed system: The maximal separation of two points
belonging to the same cycle is given by the square root of the value
of $d_{xy}$ at the upper edge of the lowest band in the distribution
of values of $d_{xy}$.

With the above method we selected exactly one point of
each periodic orbit we found.  The other points
are obtained by simply propagating the selected point with the
original map.

\subsection{Complete Sets of Orbits and Lyapunov Distributions}

In the following we first focus on the Ikeda map \cite{Ikeda79}.
It is given by
$x_{n+1}=\alpha+\beta(x_n \cos w_n - y_n \sin w_n)$,
$y_{n+1}=\beta(x_n \sin w_n + y_n \cos w_n)$,
where
$w_n = \gamma - {\delta \over {1+x_n^2+y_n^2}}$.  The attractor to be
investigated appears for $\alpha=1.0$, $\beta=0.9$, $\gamma=0.4$ and
$\delta=6.0$.
For the Ikeda map the periods $p=1,2,...,13$ 
have already been investigated in ref. \cite{Schmelcher97}. 
In order to indicate the applicability of the previously discussed algorithmic implementation
of the SD method beyond those periods we calculate complete sets of orbits for $p=14,15$.

A rough hint towards the completeness of the result 
is the convergence
of the corresponding topological entropy  $h=\lim_{p\longrightarrow
  \infty}h_p$ with $h_p=\ln n(p) / p$, $n(p)$ being the number of FPs
of all cycles of period $p$ (see table \ref{ikep1-15}).
$h$ seems to be converged fairly well to a
constant value.
Our sets of orbits for the Ikeda map, together with the corresponding results for the
H\'enon map, allow us to study the distributions of the Lyapunov exponents of the
UPOs for the two different systems (see below).

Before entering into this discussion we comment on some observations made by
applying the SD method to 
the Ikeda map. As explained 
in Sec. 2.2, each point of an UPO is stabilised by two matrices, which both
have either positive or negative
determinant. Therefore, we can group the
points of one UPO of period $p$ into the sets
\begin{eqnarray}
S_+(p)&:&\mbox{sinks, stabilised by either}
~~{\mathbf{C}}_{0+},~{\mathbf{C}}_{1+},~{\mathbf{C}}_{2+},\mbox{ or }~{\mathbf{C}}_{3+}\\
S_-(p)&:&\mbox{saddles, stabilised by either}
~~{\mathbf{C}}_{0-},~{\mathbf{C}}_{1-},~{\mathbf{C}}_{2-},\mbox{ or }~{\mathbf{C}}_{3-}
\end{eqnarray}
We found the surprising result that for all periods $p=1-14$ 
both sets $S_+(p)$ and $S_-(p)$ contain the same number of points,
$|S_+(p)|$ and $|S_-(p)|$, respectively. The values are 
$|S_\pm(p)|=1,2,4,8,11,26,36,64,121,242,419,692,1262,2256$ for $p=1-14$.
Only for $p=15$ we observe a deviation from this fact:
$|S_+(15)|=4409$, $|S_-(15)|=4379$.
At the current state we cannot 
judge whether this indicates missing orbits or represents a violation of the 
observed rule for $p=1-14$.
If we assume the validity of the equality $|S_+|=|S_-|$ as a general law, this
implies a symmetry relation between the orbits of the map: Each orbit with
a positive determinant of its stability matrix is related to
exactly one orbit with negative determinant.  In section 3, we will seek
for the most stable orbits of the Ikeda map for higher periods and we will 
find a similar pairing of orbits, i.e. two orbits, one in $S_+$ and
one in $S_-$, have nearly the same Lyapunov exponent.  This might suggest 
a symmetry of the underlying Ikeda map.
For all detected UPOs we calculate the Lyapunov exponent
$\Lambda_{orb} = \log(\vert \rho \vert)/p\label{lambdadef}$,
where $\rho$ is the largest eigenvalue of the matrix $M = M_p \cdot
\dots \cdot M_2 \cdot M_1$.  

Fig. \ref{liapdistr} shows the normalised distributions $D(\Lambda)$
of Lyapunov exponents $\Lambda$ of all orbits of order $p=24,\dots,27$
for the H\'enon map \cite{Henon76}
given by
$x_{n+1}=1.4-x_n^2+0.3 \;y_n$, $y_{n+1}=x_n$,
and for $p=12,\dots,15$ of the Ikeda map
 (primitive orbits only). For both maps the Lyapunov exponents
form a band and the distributions show a more and more pronounced peak as $p$ increases.  
The peak of the distribution appears around $\Lambda=0.5$ for the H\'enon
and around $\Lambda=0.68$ for the Ikeda map.  Around this peak, the
distribution is expected generally to be well approximated by a
Gaussian \cite{Ott93},\cite{Ellis85}.  However, globally the distributions of
both maps clearly deviate from Gaussian behaviour in some respects:
\begin{itemize}
\item Unlike symmetric distributions, both distributions exhibit
  an enhancement for small values of the Lyapunov exponent. This
  indicates that there are correlations within these orbits.
  
\item A second feature is the occurrence of peaks in the main bulk of
  the distribution. A characteristic peak of H\'enon map appears at
  $\Lambda=0.551$ and a similar spike
  is visible for low values at $\Lambda=0.435$.
  
  Probably the Lyapunov distribution for the cycles of the Ikeda map displays similar
  features, though the number of cycles for period $p=15$ is not enough to
  allow for a sufficient resolution of the distribution.
\end{itemize}
These features show that the Lyapunov distributions 
contain interesting information on the systems dynamics.

\subsection{Stability Ordering of Cycles}

Since the SD method is, by construction, changing the stability properties of a
dynamical system, it is not surprising that the value of the tuning parameter $\lambda$
is of relevance to the magnitude of the Lyapunov exponents
$\Lambda^{(j)}_{orb}$ to be detected. The
critical parameter $\lambda_{k\sigma,i}$ of a point ${\mathbf{r}}_i$ of a
cycle stabilised with ${\mathbf{C}}_{k \sigma}$ is defined as the
largest value of $\lambda$ for which both eigenvalues of the
transformed stability matrix ${\mathbf{T}}_{S_{k \sigma}}$ in eqn.
(\ref{stamattrafo}) have an absolute value less than unity, which marks the
transition from instability to stability.

As introduced in ref. \cite{Diakonos98}, an approximately monotonous relation between
$\Lambda^{(j)}_{orb}$ and the critical value
$\lambda^{(j)}_{k\sigma,i}$ can be observed. 
For the orbits of the H\'enon map, which are stabilised by a specific
${\mathbf{C}}_{k \sigma}$--matrix, this monotonous relation can be clearly seen.
However, when examining other maps, e.g. the Ikeda map,
one finds that this relation is obeyed less strictly. 
However, a slightly different concept of ordering does the job: 
We consider all points ${\mathbf{r}}^{(j)}_i$, $i=1..p$ of an
orbit $j$ of given period $p$, which in general are stabilised by
different ${\mathbf{C}}_{k \sigma}$-matrices with different critical
$\lambda^{(j)}_{k\sigma,i}$--values.  In contrast to the approach chosen in
\cite{Diakonos98} we now allow all eight ${\mathbf{C}}_{k
  \sigma}$-matrices to be used as stabilising transformations.
As explained in section 2.2, each point of the orbit $j$
is stabilised by two matrices ${\mathbf{C}}_{k \sigma}$,
${\mathbf{C}}_{k^\prime \sigma^\prime}$, with a
particular $\lambda^{(j)}_{k\sigma,i}$.  To each orbit of period $p$
there belongs a set of 2, 3 or 4 ${\mathbf{C}}_{k \sigma}$-matrices
stabilising different cycle points and a set of $2~n$
values $\lambda^{(j)}_{k\sigma,i}$.  Now we ask for the largest
$\lambda_c$ out of this set and call it the $\lambda^{(j)}_{orb}$ of
the corresponding orbit $j$. The corresponding plot for the Ikeda map
is presented in Fig. \ref{rholiap} and shows a sufficient ordering of
the stability coefficients of the detected UPOs with respect to the
corresponding critical values $\lambda^{(j)}_{orb}$.
The area in each subfigure which is shaded gray indicates the approximate position of the  
points $\Lambda^{(j)}_{orb}(\lambda^{(j)}_{orb})$ for period $p=15$. Obviously the distributions
become increasingly flatter with increasing $p$, which corresponds to a 
better stability ordering of the respective periodic orbits.
This allows us to
calculate the least unstable orbits of a map in a systematic way. We
decrease the value of $\lambda$ used in eqn. (\ref{dynsyss}) and
register the stabilised points one by one. The main difference to the
procedure in \cite{Diakonos98} is that we now have to consider the set
of all ${\mathbf{C}}_{k\sigma}$-matrices to find the $\Lambda^{(j)}$ as the
largest $\lambda^{(j)}_{k\sigma,i}$ of all points ${\mathbf{r}}_i$ of an
orbit.  
To implement these ideas, we construct a successive number of cumulative grids
$G_1,G_2,G_3,\ldots$ leading to an increasingly finer covering of the
attractor as described in Section 3.1. For our investigations of the
H\'enon and Ikeda map we used 10 grids $G_1,\ldots,G_{10}$ each with 
about 250 points. Then we perform the following steps (starting with
$\lambda=0.8$ and $i=1$):
\begin{itemize}
\item[1.]  Begin with an initial value of $\lambda$ and a grid
  $G_i$ of points.  Propagate $G_i$ 8 times with the
  stabilised systems according to (\ref{dynsyss}) for fixed $\lambda$,
  using a different ${\mathbf{C}}_{k \sigma}, k=0,...,3$, $\sigma=\pm$ each
  time.
\item[2.]  If step 1 does not yield the desired number of orbits,
  replace $\lambda\Longrightarrow r\cdot \lambda$, $r \approx 0.8$ and
  perform step 1 again.
\item[3.]  Replace $g_i\longrightarrow g_{i+1}$,
  $\lambda\longrightarrow r\cdot \lambda$ and go to step 1.
\end{itemize}
The procedure is converged if the set of the $N$ most
stable cycles resulting from the grid $H_i=G_1\cup G_2\cup...\cup
G_{i}$ remains the same compared to  the set obtained from the grid
$H_{i+1}=G_1\cup G_2\cup...\cup G_{i+1}$.  As a result of this
procedure we get a set of $N$ orbits for each grid.

For the two maps studied here we can clearly observe convergence.  In
Fig. \ref{luo} the Lyapunov exponents for the ten most stable
orbits of each period are shown for both the Ikeda and H\'enon map for
periods $p=1-36$.  These Lyapunov exponents correspond to the
lower edge of the distribution of Lyapunov exponents as given in Fig.
\ref{liapdistr}.

Two features in Fig. \ref{luo} are remarkable:
\begin{itemize}
\item The Lyapunov exponents of the least unstable orbits of a certain length $p$
 of both the H\'enon and Ikeda map are
  approximately in the same range.
This is valid for all periods $p$ considered. They
  both decrease with increasing period. This
  corresponds to a shift of the lower edge of the Lyapunov distribution 
  towards decreasing values of $\Lambda$
  with increasing period.  The inset in Fig. \ref{luo} shows the same
  Lyapunov exponents in a log--log plot. The mean of the total distribution,
i.e of all UPOs for a certain period, 
  is the average Lyapunov exponent $\bar{\Lambda}$ of the maps which is  
  approximately constant. Therefore the linear decrease in Fig. \ref{luo} implies
  that the spreading $W(p)$ of the tail of the distribution of the Lyapunov exponents
  as displayed in Fig. \ref{liapdistr} for period $p$ grows
  approximately as an algebraic function of the period, i.e.  $W(p)
  \propto p ^\eta$, $\eta>0$.
\item Both maps show exceptionally small Lyapunov exponents ( e.g.
  periods $p=13,16,18$ and $p=26,28,30$ for the H\'enon map, periods $p=19,21$
  and $p=24,27,30$ for the Ikeda map). These orbits seem to approach the main
  part of the distribution with increasing period.
\end{itemize}

\section{Summary}

This paper has two main objectives.
First it presents a novel approach towards a better understanding of the 
SD method, thereby establishing a geometric interpretation and classification, and second it
provides results of applications of the SD method to two dimensional maps.

In the first part we investigated the
stability transformations as proposed in \cite{Schmelcher97}
by introducing two classifications of the corresponding stability matrices of all FPs.
These are based on properties which change in a regular and well--defined way when the stability 
transformations are applied.
The first classification is with respect to manifolds which are invariant to the stability transformations.
The second one mirrors the change of the azimuthal angle of the flux in the
neighbourhood of the FPs.
According to these classifications of the stability matrices we can assign to 
each set of corresponding FPs a certain dynamical behaviour related to their stability properties. 
This provides new insights into the mathematical group structure which is impressed 
by the stability transformations onto the stability matrices and FPs.
Particularly this allows to determine the minimal set of transformations necessary to detect all FPs
of a given two dimensional system. We thereby learned how simple global operations on
the dynamical system change the stability properties of fixed points. 
This point of view has the advantage that it does not rely on the analytic
expressions for eigenvalues and is therefore more suited for the extension
of the method to higher dimensions.  

The second part of the present work deals with the algorithmic implementation
of the SD method, its application to the finding of UPOs in strange attractors as 
well as the evaluation and interpretation of the achieved results.
In particlar we demonstrate that even longer
cycles can be detected.
This is achieved by using a special sequence a grids of initial points for the propagation
and by an improved technique to separate distinct UPOs.
It becomes thus clear that the maximal period of orbits to be detected by 
the SD method is limited by the machine precision, not by failure of
the method itself.  
As a result of our investigations of the Ikeda map we get distributions of Lyapunov
exponents which show characteristic deviations from the first-order
approximation by a Gaussian distribution. The distributions of the
H\'enon map show similar, but distinct characteristics.  
Analysing the UPOs of the Ikeda map for the periods $p=1-15$
with respect to their stability in all different transformed systems 
suggests an underlying symmetry relation for
this map, which implies a correlation between distinct UPOs of the same period. 
As a second numerical investigation we search for the ten periodic orbits with
the smallest Lyapunov exponents for the Ikeda and H\'enon maps. Since these
orbits are the least unstable ones, it is possible to extend the
investigation up to period $p=36$ for both maps. The distributions of
the cycles as a function of the period shows a remarkably regular
overall tendency with characteristic deviations for both maps. What
is more, this part of the Lyapunov spectrum covers the section of
small values of the Lyapunov distributions which differs most from the
Gaussian approximation and might therefore provide valuable
information on the dynamical systems.

Finally we remark on very recent developments concerning the detection of
UPOs. In ref.\cite{Lai99} the SD method has been combined with the Newton-Raphson
method in order to speed up convergence in the linear neighbourhood of the
FP. Such hybrid algorithms are very desirable since they perform very efficiently
while preserving the desired global character of the SD method.

\section{Acknowledgements}

D. P. thanks the Deutsche Forschungsgesellschaft and the
Landesgraduiertenf\"orderung Baden-W\"urttemberg for financial support.
D. P. gratefully acknowledges the hospitality of Hebrew University,
Jerusalem. The hospitality of the Department of Physics (D.P. and P.S.)
of the University of Athens is appreciated.

\newpage

\vspace{2cm}
\center{\large Table Captions}
\vspace{0.5cm}

\begin{table}
\caption{Multiplication table for the matrices ${\mathbf{C}}_{k\sigma}$}
\label{cmult}
\end{table}

\begin{table}
\caption{
Transition matrices ${\mathbf{C}}_{k \sigma}$ necessary for transitions between different classes 
${\mathcal{A}}_{l\tau}$ of stability matrices of fixed points. The first three columns 
indicate the combination of classes occurring in a two dimensional chaotic dynamical system.}
\label{fptrans}
\end{table}

\begin{table}
\caption{The number of prime cycles with period $p$, the total number of cycle points of order $p$ 
and the topological entropy $h_p$
 for Ikeda map, periods $p=12,...,15$}
\label{ikep1-15}
\end{table}
\vspace{2cm}
\center{\large Figure Captions}
\vspace{0.5cm}

\begin{figure}
\caption{
Definition of the polar angles $\phi$ and $\psi_{l^\prime\tau^\prime}(\phi)$ of the displacement
${\mathbf{x}}=(\cos \phi, \sin \phi)^T$ relative to the FP at $(0,0)$ and the flux 
$F_{k\sigma}({\mathbf{x}})=(\cos \psi_{l^\prime\tau^\prime},\sin \psi_{l^\prime\tau^\prime})^T$. 
$(k \sigma)$ indicate the particular stability transformation applied according to 
eqn. (\ref{trafocontsys}). 
The indices $(l^\prime\tau^\prime)$ are given by eqn. (\ref{fptranseqn})
using $(k \sigma)$ and the indices $(l\tau)$
of the class ${\mathcal{A}}_{l\tau}$ to which the stability matrix of the original 
FP belongs.}
\label{coordsys}
\end{figure}

\begin{figure}
\caption{
a) Phase portraits of fixed points with $\det({\mathbf{B}}_{l\tau})>0$, 
	$0<\phi_{min},~\phi_{max}<\pi/2$.~~
b) and c) are the corresponding $\psi_{l \tau}(\phi)$--diagrams. 
In a) the manifolds $Z_1$, $Z_2$ (long dashed lines), eigenvectors (lines with
filled arrows) and some trajectories are shown. 
The areas shaded gray indicate the intervals of the locations of the 
eigenvectors. They correspond to the shaded boxes in the diagram c) which show these 
intervals for the four FPs discussed. }
\label{ppc1p}
\end{figure}

\begin{figure}
\caption{
  same as Fig. 2, but with $\det({\mathbf{B}}_{l
    \tau})<0$, $0<\phi_{min},~\phi_{max}<\pi/2$,~~~ }
\label{ppc1n}
\end{figure}

\begin{figure}
\caption{
   same as Fig. 2, but with $\det({\mathbf{B}}_{l
    \tau})>0$, $0<\phi_{min}<\pi/2<\phi_{max}$,~~~ }
\label{ppc2p}
\end{figure}

\begin{figure}
\caption{
    same as Fig. 2, but with $\det({\mathbf{B}}_{l
    \tau})<0$, $0<\phi_{min}<\pi/2<\phi_{max}$,~~~   }
\label{ppc2n}
\end{figure}

\begin{figure}
\caption{
  Distribution of the distances $d_{xy}$ of the orbits of the Ikeda map for period $p=15$
of the set $N_{1}$. The abscissa of the figure covers the 16076 pairs of FPs in $N_{1}$.   }
\label{dxyplot}
\end{figure}

\begin{figure}
\caption{
  Normalised distributions of the Lyapunov exponents (prime cycles only) of
  the Ikeda and H\'enon maps for various periods. }
\label{liapdistr}
\end{figure}

\begin{figure}
\caption{
  Distribution of the Lyapunov exponents of the stabilised orbits
  as a function of the critical stabilising parameter $\lambda^{(j)}_{orb}$
  for Ikeda map, periods $p=10-15$. The areas shaded gray show the approximate 
range of the distribution for $p=15$. }
\label{rholiap}
\end{figure}

\begin{figure}
\caption{
  The Lyapunov exponents of the 10 least unstable orbits of Ikeda and H\'enon map for $p=1...36$
	The inset shows the same distribution on a log--log scale.}
\label{luo}
\end{figure}

\end{document}